\documentclass{article}
\usepackage{setspace}
\usepackage{arxiv}
\usepackage{multirow}
\usepackage[table,xcdraw]{xcolor}

\usepackage{latexsym}              
\usepackage{amsmath}               
\usepackage{amssymb}               
\usepackage{amsthm}                
\usepackage{dsfont}
\usepackage[utf8]{inputenc} 
\usepackage[T1]{fontenc}    
\usepackage{hyperref}       
\usepackage{url}            
\usepackage{booktabs}       
\usepackage{amsfonts}       
\usepackage{nicefrac}       
\usepackage{microtype}      
\usepackage[affil-it]{authblk}

\usepackage[round]{natbib}
\usepackage{graphicx}
\usepackage{doi}
\usepackage{tcolorbox}
\usepackage{cleveref}

\title{Prospective Prediction of Body Mass Index Trajectories using Multi-task Gaussian Processes}

\date{23/01/2024}	

\author[1]{Arthur Leroy}
\author[2]{Varsha Gupta}
\author[3]{Mya Thway Tint}
\author[4]{Delicia Ooi Shu Qin}
\author[5]{Keith M. Godfrey}
\author[6]{Fabian Yap}
\author[7]{Leck Ngee}
\author[8]{Yung Seng Lee}
\author[9]{Johan G. Eriksson}
\author[10]{Navin Michael}
\author[11]{Mauricio A. Alvarez}
\author[12]{Dennis Wang}

\affil[----------------------------------------------------------------------------------------------------------------------------------------------------]{} 
\affil[1, 11]{Department of Computer Science, 
	The University of Manchester}
\affil[2,3,7,9,10,12]{Singapore Institute for Clinical Sciences, Agency for Science Technology and Research}
\affil[2,12]{Bioinformatics Institute, Agency for Science Technology and Research}
\affil[4]{Yong Loo Lin School of Medicine, National University of Singapore}
\affil[5]{MRC Lifecourse Epidemiology Centre 
NIHR Southampton Biomedical Research Centre \authorcr
University of Southampton and University Hospital,  
Southampton NHS Foundation Trust}
\affil[6]{KK Women's and Children's Hospital,
    Duke-NUS Medical School}
\affil[8]{Department of Pediatrics, Yong Loo Lin School of Medicine, National University of Singapore}
\affil[12]{National Heart and Lung Institute, Imperial College London} 
\affil[----------------------------------------------------------------------------------------------------------------------------------------------------]{} 

\affil[1,2]{These authors contributed equally to this work}
\affil[1,2]{Corresponding authors: \texttt{arthur.leroy.pro@gmail.com, Varsha\_gupta@sics.a-star.edu.sg}}

\renewcommand{\shorttitle}{\textit{arXiv} Template}

\hypersetup{
    pdftitle={Prospective Prediction of Body Mass Index Trajectories using Multi-task Gaussian Processes},
    pdfsubject={},
    pdfauthor={Varsha Gupta, Arthur Leroy, Dennis Wang},
    pdfkeywords={Body Mass Index, adiposity rebound, Gaussian processes, time series prediction},
    colorlinks = true,
    urlcolor = blue,
    linkcolor = orange,
    citecolor = orange
}

\begin{document}
\maketitle

\begin{abstract}
\textbf{Background:}  Clinicians often investigate the body mass index (BMI) trajectories of children to assess their growth with respect to their peers, as well as to anticipate future growth and disease risk. While retrospective modelling of BMI trajectories has been an active area of research, prospective prediction of continuous BMI trajectories from historical growth data has not been well investigated.

\textbf{Materials and Methods:} Using weight and height measurements from birth to age 10 years from a longitudinal mother-offspring cohort, we leveraged a multi-task Gaussian processes model, called \textsc{MagmaClust}, to derive probabilistic predictions for BMI trajectories over various forecasting periods. Experiments were conducted to evaluate the accuracy, sensitivity to missing values, and number of clusters. The results were compared with cubic B-spline regression and a parametric Jenss-Bayley mixed effects model. A downstream tool computing individual overweight probabilities was also proposed and evaluated.

\textbf{Results:} In all experiments, \textsc{MagmaClust} outperformed conventional models in prediction accuracy while correctly calibrating uncertainty regardless of the missing data amount (up to 90\% missing) or the forecasting period (from 2 to 8 years in the future). 
Moreover, the overweight probabilities computed from \textsc{MagmaClust}'s uncertainty quantification exhibited high specificity ($0.94$ to $0.96$) and accuracy ($0.86$ to $0.94$) in predicting the 10-year overweight status even from age 2 years. 

\textbf{Conclusion:}
\textsc{MagmaClust} provides a probabilistic non-parametric framework to prospectively predict BMI trajectories, which is robust to missing values and outperforms conventional BMI trajectory modelling approaches.
It also clusters individuals to identify typical BMI patterns (early peak, adiposity rebounds) during childhood. 
Overall, we demonstrated its potential to anticipate BMI evolution throughout childhood, allowing clinicians to implement prevention strategies. 
\end{abstract}

\keywords{Body Mass Index \and Gaussian processes \and Longitudinal forecasting \and Missing data \and Curve clustering}

\section{Introduction}
\label{sec:intro}

The increasing global prevalence of childhood obesity represents a major concern, given its strong links to comorbidities like cardiometabolic disease as well as psychopathologies \citep{wang2006worldwide,kansra2021childhood}.
Obese children have a higher prevalence of obesity-related comorbidities like insulin resistance, dysglycemia, dyslipidaemia and prehypertension/hypertension, which can track to adulthood and increase the predisposition for developing type 2 diabetes mellitus and cardiovascular disease \citep{kansra2021childhood,ford2010reduction}.
In addition, obese children also have a higher prevalence of psychopathologies like internalising (anxiety, depression, etc.) and externalising (conduct problems, attention problems, etc.) behaviours, which can persist to adulthood and result in adulthood psychiatric disorders \citep{vila2004mental, bradley2008relationship, puder2010psychological}.

Most of the current knowledge on the underlying drivers of childhood obesity, as well as its association with childhood and adulthood pathophysiology, have been drawn from studies where obesity was assessed using body mass index (BMI) based cut-offs at a single time point. 
However, BMI at a single time-point can mask significant heterogeneities in the underlying growth trajectories.
Children who are classified as obese based on the population extremes (usually 90th centile) of BMI at a single time-point do not necessarily represent a homogenous population. 
In addition, not all children who have adverse cardiometabolic or mental health are obese \citep{garcia2020normal, carsley2020association}. 
This suggests that there might be aberrant growth trajectories even within the normal weight range that increase risks of adverse cardiometabolic or mental health.
Growth is a dynamic process that is a result of the interaction between genetic growth potential and environmental influences.
Childhood growth is responsive to environmental cues (e.g. slowing growth in periods of nutritional distress and catching up in periods of nutritional abundance).
Hence, there can be considerable variability in individual growth trajectories. Individuals at the same percentile of BMI at a given time point could, therefore, have had very different growth trajectories due to differences in the histories of adverse environmental exposures over their life course.
Since maladaptive mechanistic responses to these exposures influence future disease susceptibility, children at the same percentile at any given time point could have very different accumulated risks, and obesity-related functional alterations can manifest without crossing standard obesity cut-offs. Periods of accelerated growth in childhood can result in atherosclerotic changes and increased body fat percentage, which can persist even if the child reverts to normal weight later in life \citep{tirosh2011adolescent,dulloo2006thrifty}. 
Thus, tracking and characterising longitudinal childhood growth can provide a more refined picture of predispositions for adverse adulthood health.

Parents and paediatricians are often interested in the retrospective question of how a child has grown relative to his/her peers.
This is often assessed using simplistic approaches like plotting growth assessments on childhood growth charts.
While such approaches are useful for detecting aberrant growth trends (e.g., growth faltering/stunting) which require prompt interventions, the smooth percentile curves in growth charts (generated by fitting a smooth function to the same BMI percentile across cross-sectional assessments of BMI across different ages in a reference population) provide a misleading picture of how an individual child \emph{should} grow. 
There has been a large body of literature that has focused on the problem of modelling the continuous growth curves that underlie discrete growth assessments.
This can reveal more subtle features of growth like growth velocities and growth milestones like infancy BMI peak and adiposity rebound, which have been linked to later obesity risk. 
Paediatricians are also interested in the prospective question of how a child is expected to grow in the future, given the child’s past growth history. 
While prediction of dichotomous childhood/adolescent obesity risk using early growth measurements, antenatal assessments, and early nutritional environments has been an active area of research, not many studies have systematically investigated whether continuous BMI trajectories can be prospectively forecasted from prior growth assessments. 
Knowing this would be useful for evaluating if/when a child could be expected to cross standard obesity cut-offs and for prioritising children for preventive interventions. 

In this paper, we introduce a unified probabilistic framework for simultaneously modelling, forecasting and clustering longitudinal BMI measurements during childhood while naturally handling sparse/incomplete data that leverages a multi-task Gaussian processes algorithm called \textsc{MagmaClust} \citep{leroy2023cluster}.
We demonstrate the efficiency of this framework by assessing and comparing its performance to existing growth modelling approaches using BMI data collected from a multi-ethnic Asian mother-offspring cohort. 
Furthermore, we exhibit several clinical results, such as gender patterns comparison, as well as an additional tool to compute the probability of being overweight at future ages.
 
\section{Materials and Methods}
\subsection{Training and Test Data for Growth Modeling}

Longitudinal growth data between birth and 10 years were available for 1177 children from the Growing Up in Singapore Towards healthy Outcomes (GUSTO) cohort. 
More precisely, the data collection details regarding the numbers of individuals measured at different ages are provided in \Cref{tab:number_indivs_observed}. 
Calibrated weighing scales were used for measuring weight (SECA 334 up to 18m and SECA 803 weighing scale beyond 18m). 
Recumbent length (SECA 210 mobile measuring mat) was used to compute BMI until 2 years, while standing height (SECA 213 Portable Stadiometer) was used for computing BMI beyond 2 years. 
The growth data of 1177 children were randomly split into a training set ($N = 600$) and a test set ($N = 577$).
The training set was used to train individual BMI trajectory models using 3 approaches: \textsc{MagmaClust}, multilevel Jenss-Bayley and cubic splines.
The test set was used to calculate the evaluation metrics for different experimental conditions. 
Details regarding the different growth modelling approaches and experimental conditions are detailed below.

\subsection {\textsc{MagmaClust} vs Conventional Growth Modeling Approaches}

Broadly, two classes of methods have been used for characterising longitudinal childhood BMI trends – group-based trajectory modelling (GBTM) and individual trajectory modelling. 
GBTM approaches like latent class growth analysis (LCGA) and latent class growth mixture modelling (LCGMM) have been used to identify distinct clusters of longitudinal growth trajectories, such that children within a cluster have relatively homogeneous growth trajectories \citep{nylund2007deciding,jung2008introduction,mattsson2019group}.
Grouping children based on distinctive growth patterns has been motivated by the fact that children within a cluster may share similar underlying drivers as well as future health outcomes. 
Such methods provide a probability of belonging to a growth cluster for each subject, and the mean trajectories within each cluster describe the distinctive growth patterns. 
However, such models have several limitations. Such techniques have usually been used to model relatively simple trajectory patterns (e.g. linear or quadratic trends) and face a lot of convergence issues with more complicated patterns \citep{mcneish2021improving}. 
Hence, they may not capture all possible biological trajectory patterns that exist in the population. They have been commonly used for modelling age- and sex-standardised growth metrics, which have less complex trajectory shapes than non-standardised growth metrics. 

GBTM-style approaches have been less commonly used for modelling individual growth trajectories. This has been usually performed using parametric models or nonparametric models. Parametric models use specific mathematical functions that leverage prior knowledge of the expected childhood growth trends and have biologically interpretable parameters.
A key limitation of such approaches is that while there are simple parametric forms that can model weight or height (e.g. Jenss-Bayley \citep{jenss1937mathematical} and Reed \citep{berkey1987model} models), it is difficult to capture the complex dynamics of childhood BMI in a simple parametric form. 
Instead, weight and height have to be modelled separately, which can then be used to estimate the BMI curve \citep{carles2016novel}. Hence, errors in the individual weight and height models can propagate to the estimated BMI model. 
Alternatively, BMI can be directly modelled using flexible functions that can model arbitrary shapes like fractional polynomials and splines \citep{tilling2014modelling}.
In these models, the estimated parameters have no biological meaning. However, such approaches are optimised for interpolating between observed growth measurements and may have poor performance extrapolating growth trends outside the observed window.

In the current work, we proposed to use a recent multi-task Gaussian processes algorithm called \textsc{MagmaClust} \citep{leroy2023cluster}, which has been previously used for time series forecasting for growth modelling. 
In contrast to the previous growth models, Gaussian processes-based methods offer a probabilistic non-parametric framework by defining a prior distribution over functions, allowing us to capture complex non-linear relationships while accounting for uncertainty.
\textsc{MagmaClust} performs functional curve clustering as well as prediction of individual trajectories within the same model, obviating the need for separate approaches for growth clustering and individual trajectory modelling. 
Thus, it represents an advancement over traditional GBTM and individual trajectory modelling approaches.

Formally, a Gaussian process (GP) is a random process over functions (or curves) that is characterised by a specific mean and covariance function.
Intuitively, GPs generalise traditional multivariate Gaussian distributions as any evaluation of a GP at a finite number of points is a multivariate Gaussian, parametrised by the corresponding mean vector and covariance matrix (see the monograph \cite{RasmussenGaussianProcessesMachine2005} for detailed explanations).  
Observed individual growth trajectories can be visualised as specific instantiations of different Gaussian processes.  
A naïve approach would be to model each individual trajectory using separate, independent GPs. 
However, this ignores the structure that exists in the data since the individual trajectories all represent the same underlying growth process and the potential for improving learning by sharing information across the different individual trajectories. 

In an earlier iteration of the framework (called \textsc{Magma} in \cite{leroy2022magma}), this information sharing was achieved by expressing the trajectory of each individual as a sum of a common mean GP shared by all individuals, and an individual-specific GP.
\textsc{MagmaClust} advances this approach by allowing the simultaneous clustering of growth patterns and defining a  common mean GP for each cluster specifically.
Similar to LCGMM, \textsc{MagmaClust} is a mixture model that returns the membership probability of different clusters for each individual (individuals can have non-zero probabilities of belonging to multiple clusters). 
Moreover, the final prediction of an individual growth curve is expressed as a GP mixture of all cluster-specific predictions, weighted by the adequate membership probabilities. 
The sharing of longitudinal growth information across multiple individuals and the allowance for clustering of growth trajectories offer more accurate individual predictions while accounting for uncertainty thanks to the probabilistic nature of GPs. 
The method can accommodate arbitrary trajectory shapes, naturally deals with irregular measurements, and has been designed to provide robust predictions even with missing values. 
As a summary, let us provide in \Cref{fig:flowchart} a flowchart of the main steps of the \textsc{MagmaClust} algorithm as described in its original article. 

\subsection{Cubic B-splines (fixed effects only)}

We also conducted the experiments using a standard splines approach as a baseline. 
More specifically, we defined our BMI trajectories as a decomposition of cubic B-splines (see \cite{de1972calculating, de1978practical} for technical details). 
Each child was treated individually by fitting an independent B-spline decomposition on its data points.
The smoothing computations were performed thanks to the \emph{smooth.spline} function of the \emph{stat} R package. 
While utilising cubic B-splines is a standard choice to obtain a smooth and flexible fit for functions, it induces a minimal requirement of 4 data points to be computed. 
Therefore, some experiments involving missing data could lead to numerical errors when the amount of observed data was too low for an individual. 
We specifically referred to those cases as \emph{failed computations} in \Cref{tab:missing_data_ratio}.

\subsection{Parameteric Jenss-Bayley with random effects}

The Jenss-Bayley parametric growth model was originally proposed for modelling weight and height between birth and 6-8 years \cite{jenss1937mathematical}.
We used a modified and parameterised form of the
Jenss-Bayley model \citep{botton2008postnatal, botton2014postnatal} for modelling both weight and height/length that includes an additional quadratic term that can account for growth during puberty and is suitable for modelling growth up to age 12 years. The modified Jenss-Bayley weight and height models were hierarchically fitted with a non-linear mixed effect model using the SAEMIX package in R.  Children with at least two measurements of weight and height were included in the analysis (1106 children). The individually fitted weight and height trajectories were subsequently used for calculating the BMI at different time points.

\subsection{Evaluation metrics}
\label{sec:metrics}
For clarity, let us recall that $N$ denotes the number of individuals, $T_i$ the number of time points observed for the $i$-th individual, and $K$ is the number of clusters, whereas $y_{obs}$ and $y_{pred}$ represent the functions of observed and predicted BMI, respectively.
Formally, we define the Mean Squared Error (MSE) in the subsequent experiments as follows:
\begin{equation*}
	\dfrac{1}{N T_i} \sum\limits_{i = 1}^{N}  \sum\limits_{t = 1}^{T_i} (y_i^{obs}(t) - y_i^{pred}(t))^2.
\end{equation*}
Moreover, an additional measure of uncertainty quantification, introduced in \cite{leroy2023cluster}, is used to evaluate whether the observations belong to the predicted credible intervals as expected.
Namely, the weighted $CI_{95}$ coverage ($WCIC_{95}$) is defined as:
\begin{equation*}
	100 \times \dfrac{1}{N} \sum\limits_{i = 1}^{N} \sum\limits_{K = 1}^{K} \tau_{ik} \ \mathds{1}_{ \{ y_i^{obs} \in \ CI_{95}^{k} \}},
\end{equation*}
\noindent where $CI_{95}^{k}$ represents the 95\% credible interval computed for the $k$-th cluster, and $\tau_{ik}$ corresponds to the probability for the $i$-th individual to belong to the $k$-th cluster.
When interpreting this metric, the closer to the theoretical value of 95\%, the better. 

\subsection{BMI prediction experiments}

We first illustrate the operation and advantages of the \textsc{MagmaClust} framework and then extensively evaluate its performances, particularly in comparison with natural competitors, such as B-splines or Jenss-Bayley. 
Throughout the experiments, we split the dataset of BMI measurements into a \emph{training} (600 individuals) and a \emph{testing} set (577 individuals).
The evaluation of performances is assessed through various settings involving \emph{testing} individuals, for which we compute metrics, as introduced in \Cref{sec:metrics}, to measure prediction errors (MSE) and quality of uncertainty quantification ($WCIC_{95}$).
Each following experimental setting focuses on a particular aspect (e.g. robustness to missing data, etc.) and is reported within a dedicated subsection.

The overall problem of BMI values prediction is considered through two separate tasks: missing data reconstruction and forecasting. 
The former is intuitively seen as simpler since the missing points are distributed within the observed range of values, whereas the latter generally leads to higher uncertainty in long-term predictions (as summarised by the famous maxim: \emph{It's difficult to make predictions, especially about the future}).
Despite being disparate challenges, those tasks only differ, from a mathematical point of view, in the location of points we aim to predict.
Unless otherwise stated in the sequel, missing data reconstruction tasks consist of randomly removing 50\% of the observed points for each individual and using the remaining 50\% to predict the BMI time series between 0 and 10 years.
In contrast, in forecasting tasks, we retain all points observed after 6 years to be testing points and only use data before 6 years to predict BMI time series between 0 and 10 years. 
Then, performance metrics are computed at the testing points to measure errors between observed and predicted values. 
Throughout, those metrics are reported with the "\emph{mean (sd)}" format, where the mean and standard deviation are computed across all individuals.
\begin{figure}
	\centering
        \includegraphics[width = \textwidth]{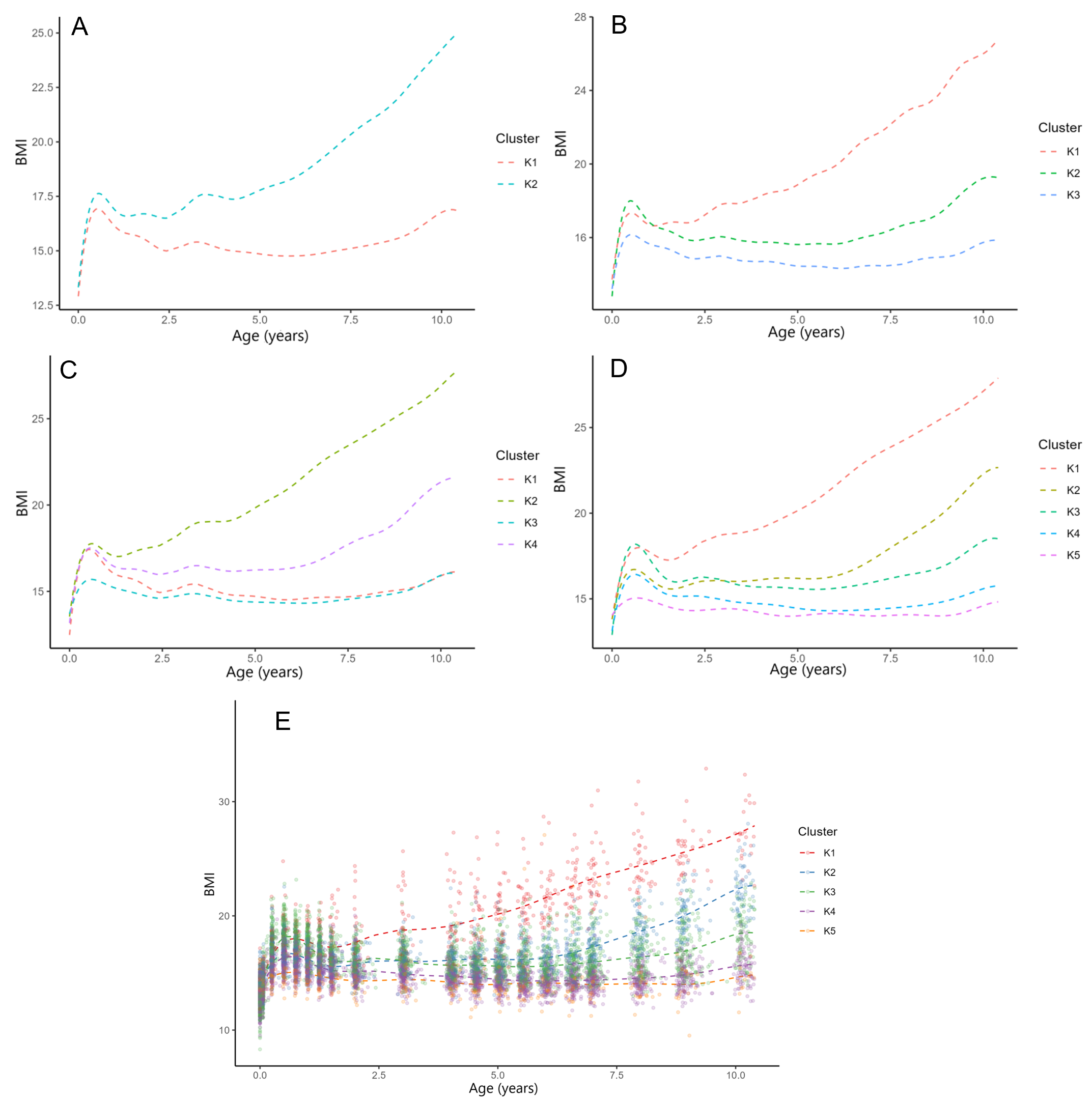}
	\caption{ \textbf{(A-D):} Cluster-specific mean BMI curves with increasing number of clusters: \textbf{A:} K=2 ,  \textbf{B:} K=3, \textbf{C:} K=4, and \textbf{D:} K=5. \textbf{(E):} Mean BMI curves associated with K = 5 overlaid on observations from the training data set, coloured according to their most probable cluster.}
	\label{fig:magma_cluster}
\end{figure}

\section{Results}

\subsection{MagmaClust: Number of clusters and Cluster-specific mean processes}

The number of clusters, $K$,  is a user-defined parameter in the MagmaClust approach. The cluster-specific mean BMI trends with increasing number of clusters are illustrated in \Cref{fig:magma_cluster}(A-D). 
We also repeated the MagmaClust approach with increasing cluster numbers (up to 10 clusters). However, for $K \geq 6$, the additional clusters generated are empty or contain very few (generally only one) individuals.
This behaviour, with increasing cluster number, suggested that 5 clusters are enough to capture the main trends present in the current dataset. 
We note that choosing an adequate value of $K$ is more of a practitioner's trade-off between interpretability and complexity. 
As we increase the number of clusters, we can observe that the global trend tends to split into more subtle sub-patterns. 
For instance, while the upper cluster, growing towards BMI values of 26 at 10 years, remains roughly similar, the other cluster (when $K=2$) seems to split into more specific sub-clusters as $K$ increases.
Even if we observe a peak in BMI values for all groups around 9 months, the intensity of this peak and the subsequent evolution during childhood can vary through patterns that seem well captured by the algorithm when $K=5$. 
Although those mean processes are mainly computed in \textsc{MagmaClust} for technical reasons (to transfer knowledge across individuals and provide more accurate predictions), we believe that identifying these mean trends already constitutes a relevant outcome for practitioners studying the evolution of BMI over time.

\Cref{fig:magma_cluster}(E) depicts the mean curves associated with each of the 5 clusters, overlaid on top of the complete dataset used during training.
This graph highlights how the algorithm identifies underlying patterns to simultaneously optimise parameters to estimate mean processes accurately and attribute membership in each cluster for all individuals. 
It is interesting to notice that, although underlying characteristic patterns can be captured through each cluster's mean process, a continuum of data points exists in between.
This critical aspect is taken into account in mixture models like \textsc{MagmaClust}, as no individual strictly belongs to one cluster.
Instead, a weight corresponding to the membership probability is computed for each cluster. As the sum of all weights equals 1, we may adopt a probabilistic interpretation that each individual belongs to a \emph{mixture of clusters}.
In \Cref{fig:magma_cluster}(E), we coloured individuals according to their most probable clusters, mainly for visual convenience (readers interested in technical details about the mixture model and its interpretation can refer to \cite{leroy2023cluster}, Sections 4 and 6.1).

\subsection{Missing Data Reconstruction}

A recurrent concern in biological or medical studies often comes from the presence of missing data in measurements.
Methods reconstructing curves from sets of points are, by definition, providing values at all unobserved locations. 
From a mathematical point of view, missing data in time series are merely unobserved points, like any others. 
Therefore, we can merely treat missing data problems as a particular case of general functional regression. 
As a motivating example, we first illustrate the advantages of the \textsc{MagmaClust} framework compared to standard splines regression and Jenss-Bayley's methods for handling missing data reconstruction. 
\begin{figure}
	\centering
        \includegraphics[width = \textwidth]{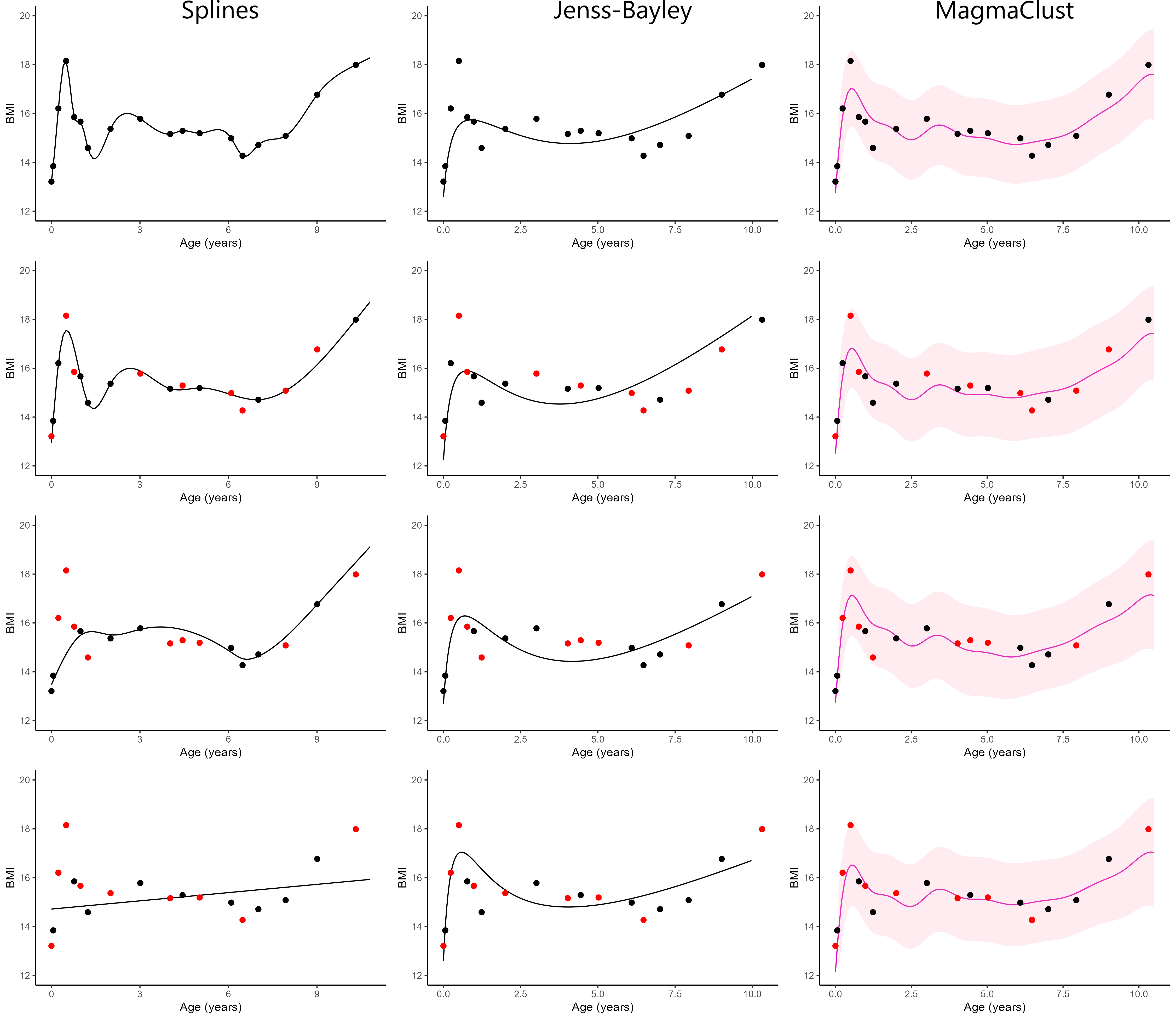}
	\caption{Depending on which points are missing, B-splines (\textbf{left}) modelling can lead to large differences in reconstruction, while Jenss-Bayley (\textbf{middle}) is more robust, and \textsc{MagmaClust} (\textbf{right}) provides robustness, an accurate fit and uncertainty quantification.}
	\label{fig:splines_comparison}
\end{figure}
In \Cref{fig:splines_comparison}, half of the points (in red) are randomly removed from the observed data of one individual to emulate the missing data paradigm. 
The remaining observations (in black) are used to reconstruct the time series thanks to B-splines (left panel) and \textsc{MagmaClust} (right panel). 
In each row, a different subset of red points is removed to compute the prediction curves with both methods. 
In the left panel, one can observe that the B-splines fitted curves can dramatically vary depending on which points are missing. 
In the most pathological case (bottom left graph), we even obtain a straight line, largely under-fitting the original signal.  
Practitioners using them regularly are well aware that this behaviour is commonplace, as splines are known to be particularly sensitive to the lack of data and boundary conditions \citep{de1978practical}.
On the other hand, the right panel shows that \textsc{MagmaClust} predictions remain remarkably robust regardless of the observed subset. When it comes to making predictions, as for any GP-based method, the MagmaClust procedure involves the derivation of a posterior Gaussian distribution in closed form, which is characterised by its mean and variance parameters. As this distribution can be computed for any unobserved time point, we generally represent results as a mean curve, along with a credible interval. Even with different patterns of missingness, MagmaClust was able to recover the correct mean trend while also providing an accurate uncertainty quantification with the associated 95\% credible interval (pink region).
This ability to capture uncertainty for the predicted curve is another major advantage of probabilistic methods compared to frequentist approaches like splines. 
As a reminder, the 95\% credible interval ($CI_{95}$) corresponds, at any instant, to the range of values in which the prediction has a probability of 0.95 to belong (an intuitive interpretation that one should not confound with the \emph{confidence} interval one, which is often misunderstood \citep{hoekstra2014robust}).

Although compelling in this visual example for one individual, we performed additional experiments to rigorously test the performance of the \textsc{MagmaClust} approach over legacy methods with hundreds of individuals as follows. 
Firstly, to emulate the presence of missing data in the time series, we randomly removed 50\% of the observations in all 577 individuals in the test set and used the remaining observations to compute predictions. 
\begin{table}
\centering
\begin{tabular}{c|cc|}
\cline{2-3}
  & MSE   & $WCIC_{95}$    \\ \hline
\multicolumn{1}{|c|}{\textsc{MagmaClust} 2 clusters}  & 1.55 (5.88)   & 92.58 (11.23)  \\
\multicolumn{1}{|c|}{\textsc{MagmaClust} 3 clusters}  & 1.68 (6.93)   & 90.92 (12.16)  \\
\multicolumn{1}{|c|}{\textsc{MagmaClust} 4 clusters}  & 1.64 (6.79)   & 92.56 (11.69)  \\
\multicolumn{1}{|c|}{\textsc{MagmaClust} 5 clusters}  & 1.69 (6.47)   & 91.07 (12.22)  \\
\multicolumn{1}{|c|}{Jenss-Bayley}  & 2.22 (4.31) & Not applicable \\
\multicolumn{1}{|c|}{Splines}    & 8.11 (258.17) & Not applicable \\ \hline
\end{tabular}
\caption{Average (sd) values of MSE, $WCIC_{95}$ in missing data reconstruction for 577 testing individuals when applying \textsc{MagmaClust}  for different numbers of clusters, Jenss-Bayley, and Splines.}
\label{tab:missing_all_indivs}
\end{table}
The missing data reconstruction errors for \textsc{MagmaClust} (from 2 to 5 clusters), Jenss-Bayley and splines are reported in \Cref{tab:missing_all_indivs}.
We observed low mean square errors (MSE) for \textsc{MagmaClust}, no matter the number of clusters.
The MSE of the 5-cluster \textsc{MagmaClust} model was 24\% lower than the Jens-Bayley model and 79\% lower than the spline model.
Overall, the methods sharing information across individuals (\textsc{MagmaClust} and multi-level Jens-Bayley) exhibited better performance at missing data reconstruction than methods merely relying on individual data.
\Cref{fig:error_missing_uncertainty} (left panel) depicts the predicted curve for a random individual with 50\% missing data obtained by the 5-cluster \textsc{MagmaClust} model overlaid on the cluster-specific mean curves.
Missing data reconstruction is well handled by \textsc{MagmaClust} as it can leverage data distributed over the whole time interval and, roughly speaking, 'fill the blanks' in between. 
The narrow band of probable values highlights the relatively low uncertainty associated with predictions in this context. 
The uncertainty is also well calibrated as the $WCIC_{95}$ metrics are close to the theoretical value of 95\%.
Another illustration of the overall performances with 5 clusters is provided in \Cref{fig:error_missing_uncertainty} (right panel).
This graph displays the error between predicted and observed BMI values, sorted by increasing variance. 
It can be noticed that the pink region, accounting for the expected uncertainty, adequately recovers the range of most errors, as desired.

Whereas the previous experiment assumed a 50\% proportion of missing values somewhat arbitrarily, this ratio can vary highly depending on the context in real-life applications.
Hence, we compared the reconstruction performance of the 5-cluster \textsc{MagmaClust} model with Jens-Bayeley and cubic splines with the proportion of missing data ranging from 10\% to 90\% in \Cref{tab:missing_data_ratio}. 
\begin{table}
\centering
\begin{tabular}{c|cc|c|cc|}
\cline{2-6}   & \multicolumn{2}{c|}{\textsc{MagmaClust}} & Jenss-Bayley & \multicolumn{2}{c|}{Splines}      \\ \cline{1-1}
\multicolumn{1}{|c|}{Missing data ratio} & MSE   & $WCIC_{95}$  & MSE   & MSE  & Failed computations \\ \hline
\multicolumn{1}{|c|}{10\%}    & 0.90 (2.36)   & 90.90 (28.8)    & 0.94 (1.99) & 1.86 (4.78) & 3.1\%               \\
\multicolumn{1}{|c|}{25\%}    & 1.39 (3.17)   & 91.60 (15.84)   & 1.55 (1.96) & 2.60 (8.61) & 2.5\%               \\
\multicolumn{1}{|c|}{50\%} & 1.71 (3.16)   & 91.38 (12.64)   & 2.63(2.98)  & 3.57 (21.5) & 5.2\%               \\
\multicolumn{1}{|c|}{75\%}    & 2.00 (2.97)   & 93.00 (10.84)   & 3.29(4.44)  & 5.14 (15.5) & 68.5\%              \\
\multicolumn{1}{|c|}{90\%}   & 2.84 (8.74)   & 95.06 (9.26)    & 8.06 (9.49) & /           & 100\%               \\ \hline
\end{tabular}
\caption{Average (sd) values of MSE, $WCIC_{95}$ in missing data reconstruction with an increasing percentage of missing data, for 577 individuals when applying \textsc{MagmaClust} with 5 clusters, Jenss-Bayley, and Splines}
\label{tab:missing_data_ratio}
\end{table}
As before, the remaining observations (sometimes only one data point is left in 90\% missing settings) are used to predict the removed values. 
One can observe how the mean squared errors of \textsc{MagmaClust} increase only slightly when we increment the missing data ratio while preserving a remarkably accurate quantification of uncertainty (close to the theoretically expected 95\%).
Those results highlight the usefulness of cluster-specific mean processes, identifying the most probable trajectories even from a handful of data points and providing a reliable estimation for the unobserved locations. 
Conversely, while cubic splines and Jenss-Bayley remain reasonably efficient for low proportions (below 50\%), they typically struggle as the missing data ratio increases. 

Even more concerning than the high error rate, the occurrence of pathological cases leading to computational errors for Splines, existing even for low ratios, start to explode above 50\% of missing points. 
(For instance, the splines estimation failed for 68.5\% of the 577 individuals when removing three-quarters of observed data, and no results could be obtained when increasing the missing proportion to  90\% (typically, the classical \emph{cubic} splines require a minimum of 4 points to be estimated).


\subsection{Forecasting BMI Growth Curves in Childhood}

\begin{table}
\centering
\begin{tabular}{c|cc|c|c|}
\cline{2-5}
& \multicolumn{2}{c|}{\textsc{MagmaClust}} & Jenss-Bayley & Splines          \\ \cline{1-1}
\multicolumn{1}{|c|}{Forecasting}        & MSE           & $WCIC_{95}$     & MSE         & MSE              \\ \hline
\multicolumn{1}{|c|}{from 2 to 10 years} & 4.11 (12.77)  & 95.21 (21.361)  &       13.39(22.57)      & 352.47 (1386.57) \\
\multicolumn{1}{|c|}{from 3 to 10 years} & 3.94 (12.65)  & 94.55 (22.71)   &    12.31(19.03)         & 156.73 (409.61)  \\
\multicolumn{1}{|c|}{from 4 to 10 years} & 3.11 (8.83)   & 95.11 (21.58)   &          8.69(13.27)   & 52.10 (157.69)   \\
\multicolumn{1}{|c|}{from 5 to 10 years} & 2.81 (8.60)   & 94.47 (22.85)   &          5.46(8.37)   & 26.10 (95.26)    \\
\multicolumn{1}{|c|}{from 6 to 10 years} & 2.55 (7.96)   & 94.46 (22.88)   & 4.18(7.59)  & 23.06 (88.73)    \\ \hline
\end{tabular}
\newline
\caption{Average (sd) values of MSE and $WCIC_{95}$ in forecasting using an increasing number of observed early data points for 577 testing individuals when applying \textsc{MagmaClust} with 5 clusters, Jenss-Bayley, and Splines.}
\label{tab:forecast_all_indivs}
\end{table}
When working with time series, a classical motivation, generally referred to as \emph{forecasting}, consists in extrapolating current observations into the future. 
In our experiments, the problem is roughly similar to missing data reconstruction, except that the points are removed at the end of the observation interval (more specifically, all points after 2, 3, 4, 5 or 6 years are used for testing purposes).  We report the forecasting performance of \textsc{MagmaClust}, splines and Jenss-Bayley models for all individuals in the test set, for forecasting periods of 2 to 10y, 3 to 10y, 4 to 10y, 5 to 10y and 6 to 10y, in \Cref{tab:forecast_all_indivs}.
Note that the errors for the models in \Cref{tab:forecast_all_indivs} are higher than those of the models in \Cref{tab:missing_data_ratio}. 
This behaviour is expected as forecasting is a more challenging mathematical problem than missing data reconstruction. 
The MSE from Jenss-Bayley was 1.62 times higher than \textsc{MagmaClust} when forecasting from 6 to 10 years.
This error ratio between the two methods increases to 3.26 when forecasting is performed using data from birth to 2 years only.
Also, it should be noted that these experiments could only be performed for Jenss-Bayley when 2 data points were collected during the fitting period, while \textsc{MagmaClust} still runs with only one. 
The performance of splines was much worse, with 9-fold higher MSE for 6 to 10y and 86-fold higher MSE for 2 to 10y when compared to \textsc{MagmaClust}. 
These findings highlight the unsuitability of splines for the purpose of prospective forecasting due to large errors due to boundary issues like local linear extrapolations. 
Once again, it seems that transferring information across individuals to identify credible trajectory evolution is particularly efficient for this forecasting problem. \textsc{MagmaClust}  provides remarkably accurate predictions, outperforming Jenss-Bayley for all prediction ranges. 
More importantly, uncertainty quantification appears nearly perfect as the $WCIC_{95}$ remains really close to the expected $95\%$ value in all settings.
Overall, by identifying relevant patterns from early BMI measurements, \textsc{MagmaClust} demonstrates its ability to forecast probable trajectories accurately, even several years in the future.

\begin{figure}
	\centering
        \includegraphics[width = \textwidth]{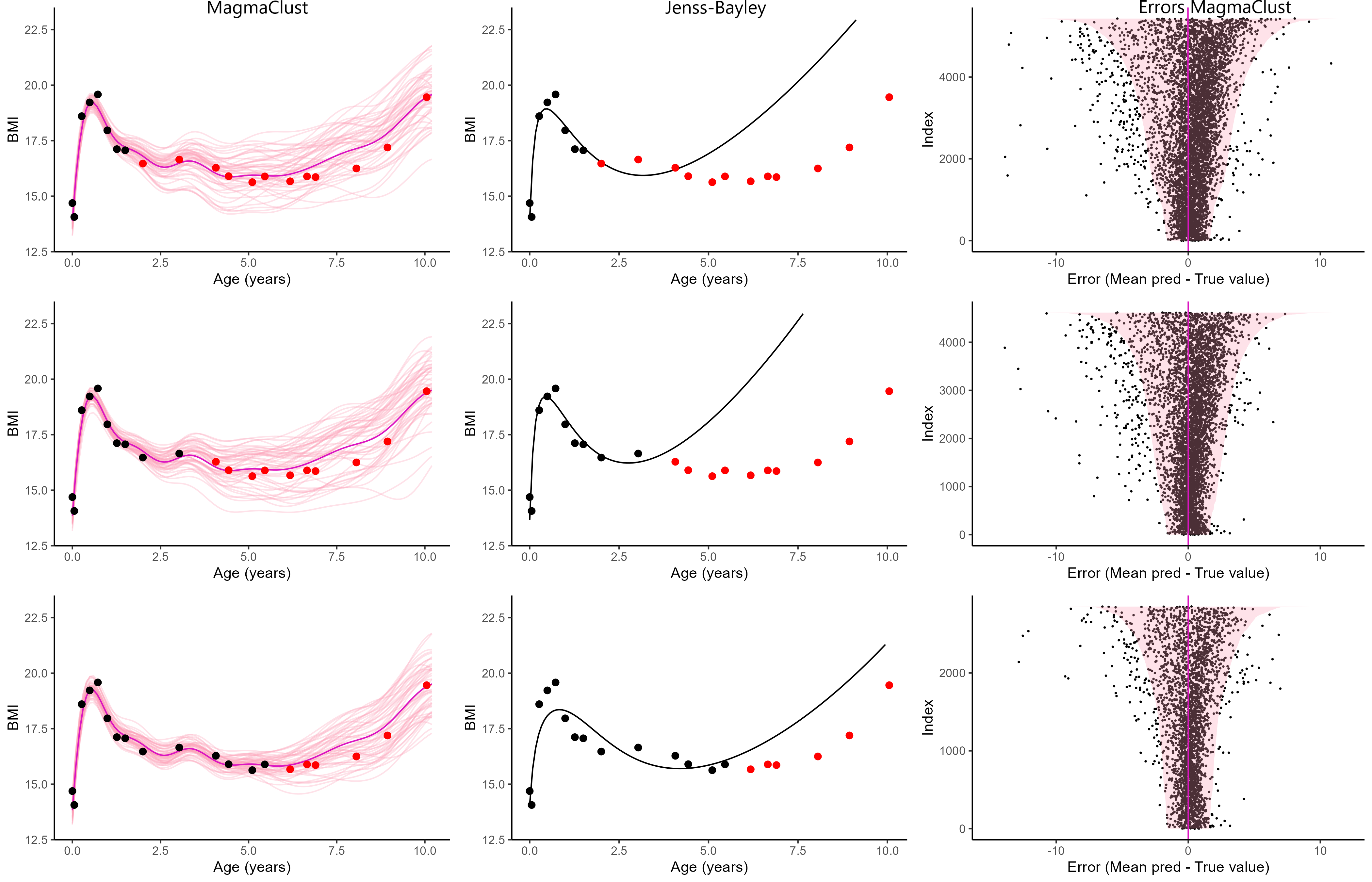}
	\caption{\textbf{Left:} Illustration of \textsc{MagmaClust} forecasts for a random illustrative individual, observed until 2 years (top), 4 years (middle), and 6 years (bottom). Observed points are in black, while testing points are in red. The purple curve represents the mean prediction, whereas the pink curves correspond to 50 samples drawn from the posterior to highlight the prediction uncertainty; \textbf{Left:} Similar illustration for Jenss-Bayley's forecasts; \textbf{Right:} Overall uncertainty quantification of errors for \textsc{MagmaClust}'s forecasts across all individuals and all testing points. For each individual index (sorted by decreasing uncertainty on the y-axis), the 0 purple line on the x-axis represents its posterior mean; the pink region corresponds to the associate 95\% credible interval; and the black dot is the absolute error to the true value}
	\label{fig:error_pred_uncertainty}
\end{figure}
To provide more visual intuition, we displayed prediction results of \textsc{MagmaClust} and Jenss-Bayley for one individual for 3 prediction ranges (2 to 10, 4 to 10 and 6 to 10 years) in \Cref{fig:error_pred_uncertainty}.
Note that, for a multi-task Gaussian model like \textsc{MagmaClust}, the predictive distribution is not strictly Gaussian but a mixture of GPs (i.e. a linear combination of Gaussian distributions, each associated with one cluster, for which the weight corresponds to its cluster's membership probability). 
This means that the resulting distribution may not be unimodal anymore.
Therefore, predictions are generally displayed through sample curves drawn from the posterior distribution, as in \Cref{fig:error_missing_uncertainty} and \Cref{fig:error_pred_uncertainty} (left panels). 
This representation is actually more informative in all cases and should be favoured when illustrating Gaussian process predictions. 
It highlights the varying uncertainty over time by representing the multiplicity of \emph{probable} trajectories considering our current knowledge. 
In the absence of observed data (black dots), we can observe that Jenss-Bayley predictions quickly diverge from testing values (red dots). 
On the contrary, \textsc{MagmaClust} forecasts provide an accurate mean trend in all cases, and observing more data seems to only narrow the range of probable trajectories. 

The right panel of \Cref{fig:error_pred_uncertainty} displays the forecasting errors (black dots) of \textsc{MagmaClust} while highlighting the 95\% credible intervals coverage (pink area) associated with its predictions. 
One can notice that the prediction errors were well anticipated by the method as the vast majority (around 95\%, as theoretically expected) of the black dots overlap the pink credible interval.
This visual evidence corroborates the results from \Cref{tab:forecast_all_indivs} regarding the correct calibration of uncertainty quantification.

\subsection {Genders comparison}

\begin{figure}
	\centering
        \includegraphics[width = 0.7\textwidth]{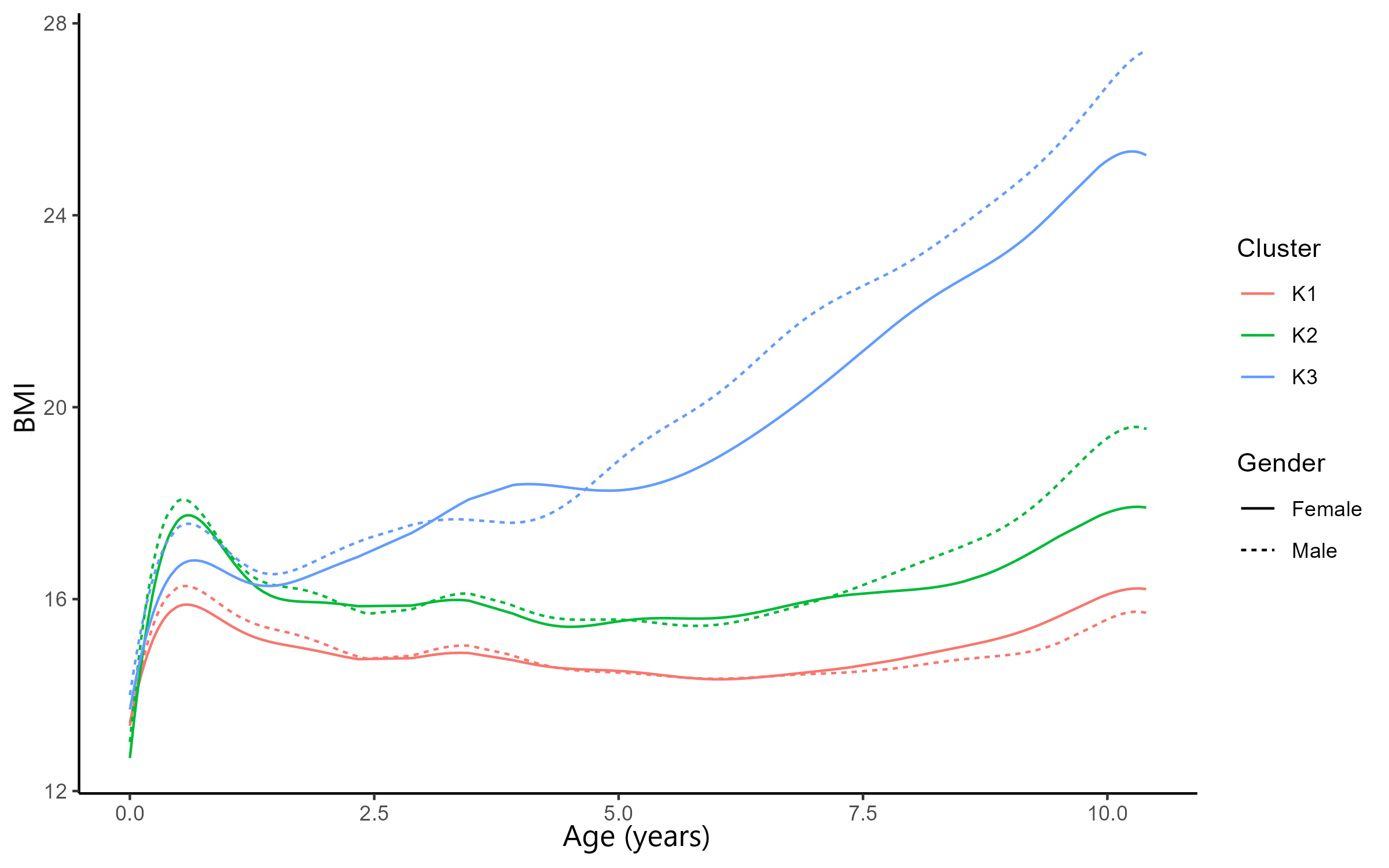}
	\caption{Mean curves comparison according to the gender for K = 3 clusters.}
	\label{fig:gender_comparison}
\end{figure}
To pursue the matter of forecasting, let us report incidental results that appeared rather unexpectedly from previous experiments. 
Whereas we identified that both boys and girls presented roughly similar BMI evolution patterns between birth and 10 years, as displayed in \Cref{fig:gender_comparison} (we represented 3 clusters per gender for clarity), it appeared on \Cref{tab:forecast_by_gender} that errors were consistently higher when predicting male individuals, for all methods and settings. 
Intuitively, we might attribute this property to higher variability in boys' BMI values during childhood.
Although this hypothesis would necessitate further investigations beyond the present paper's ambitions, we can mention that similar conclusions were previously reported in \cite{boyer2015childhood}.
Whilst we described the trends as roughly similar for both genders in \Cref{fig:gender_comparison}, we shall still notice an interesting difference occurring during the early BMI peak around 9 months. 
In all clusters, the value of the peak seemed to be slightly higher for boys (dashed lines) than for girls (plain lines), although the location of this peak remained apparently synchronised in corresponding groups. 
Following this noticeable gap, the gender trends tend to overlap mostly before diverging slightly after 5 years. 
\begin{table}
\centering
\begin{tabular}{cc|cc|c|c|}
\cline{3-6}  &  & \multicolumn{2}{c|}{MagmaClust} & Jenss-Bayley & Splines \\ 
&        & MSE    & $WCIC_{95}$     & MSE    & MSE  \\ \hline
\multicolumn{1}{|c}{\multirow{2}{*}{Missing data}} & Female & \multicolumn{1}{l}{1.21 (1.50)} & \multicolumn{1}{l|}{93.05 (9.76)} & 2.36(2.34)  & 2.79 (7.90)  \\
\multicolumn{1}{|c}{}  & Male   & 1.93 (3.60)  & 91.61 (12.60)  & 2.78(2.92)  & 5.16 (37.0)  \\ \hline
\multicolumn{1}{|c}{\multirow{2}{*}{Forecasting}}  & Female & 1.89 (4.65)   & 92.65 (14.67) & 3.06 (4.78) & 20.2 (89.40) \\
\multicolumn{1}{|c}{}   & Male   & 3.11 (4.80)   & 92.06 (16.58)  & 3.97 (7.38) & 28.9 (139.0) \\ \hline
\end{tabular}
\caption{Average (sd) values of MSE, $WCIC_{95}$ in forecasting for 243 female and 277 male testing individuals using \textsc{MagmaClust} with 3 clusters.}
\label{tab:forecast_by_gender}
\end{table}

\subsection{Predict Overweight Status during Childhood}

\textsc{MagmaClust} can also be used to provide a more practical tool to help paediatricians in their longitudinal follow-up of children's growth. 
A classical medical concern regarding BMI is the divergence from typical and healthy patterns, leading to overweight or underweight. 
Therefore, we leveraged our previous BMI forecasts to derive a measure of the \emph{probability of being overweight at 10 years} (the choice of overweight rather than underweight and the age of 10 years is arbitrary; the tool can be easily adapted for any weight thresholds and age between 0 to 10y). 
In terms of convention, let us note that the overweight threshold at 10 years differs according to sex: $BMI > 22$ for girls and $BMI > 22.8$ for boys.
With these values in mind, we can now utilise the samples of probable trajectories provided by \textsc{MagmaClust} to count the proportion of curves exceeding the threshold at 10 years. 
To provide an accurate quantification, we sampled for each child 100,000 trajectories from their predictive distribution and computed the ratio of those leading to overweight values among the total number of samples (100,000 in our case).
\begin{figure}
	\centering
        \includegraphics[width = 0.49\textwidth]{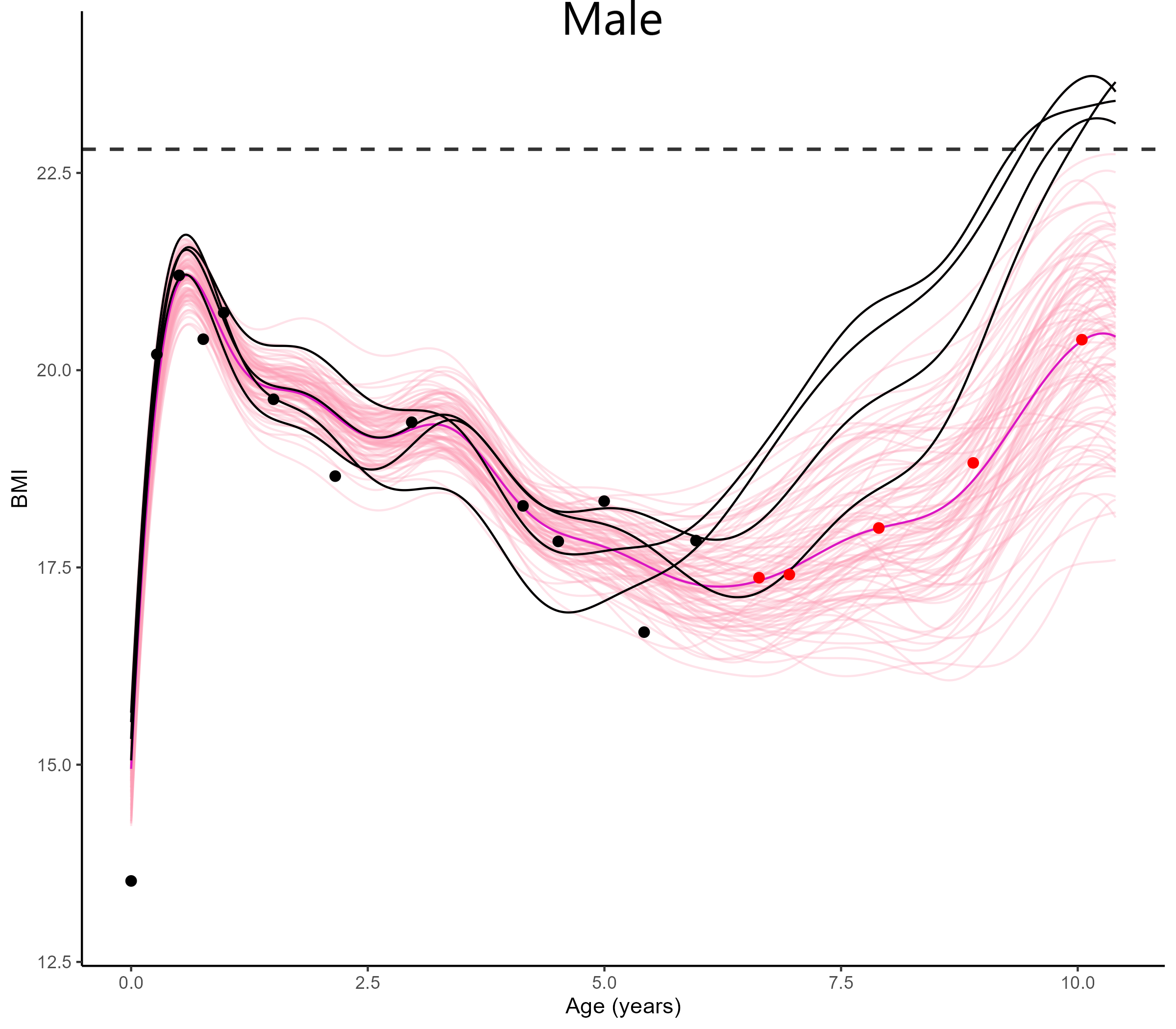}
        \includegraphics[width = 0.49\textwidth]{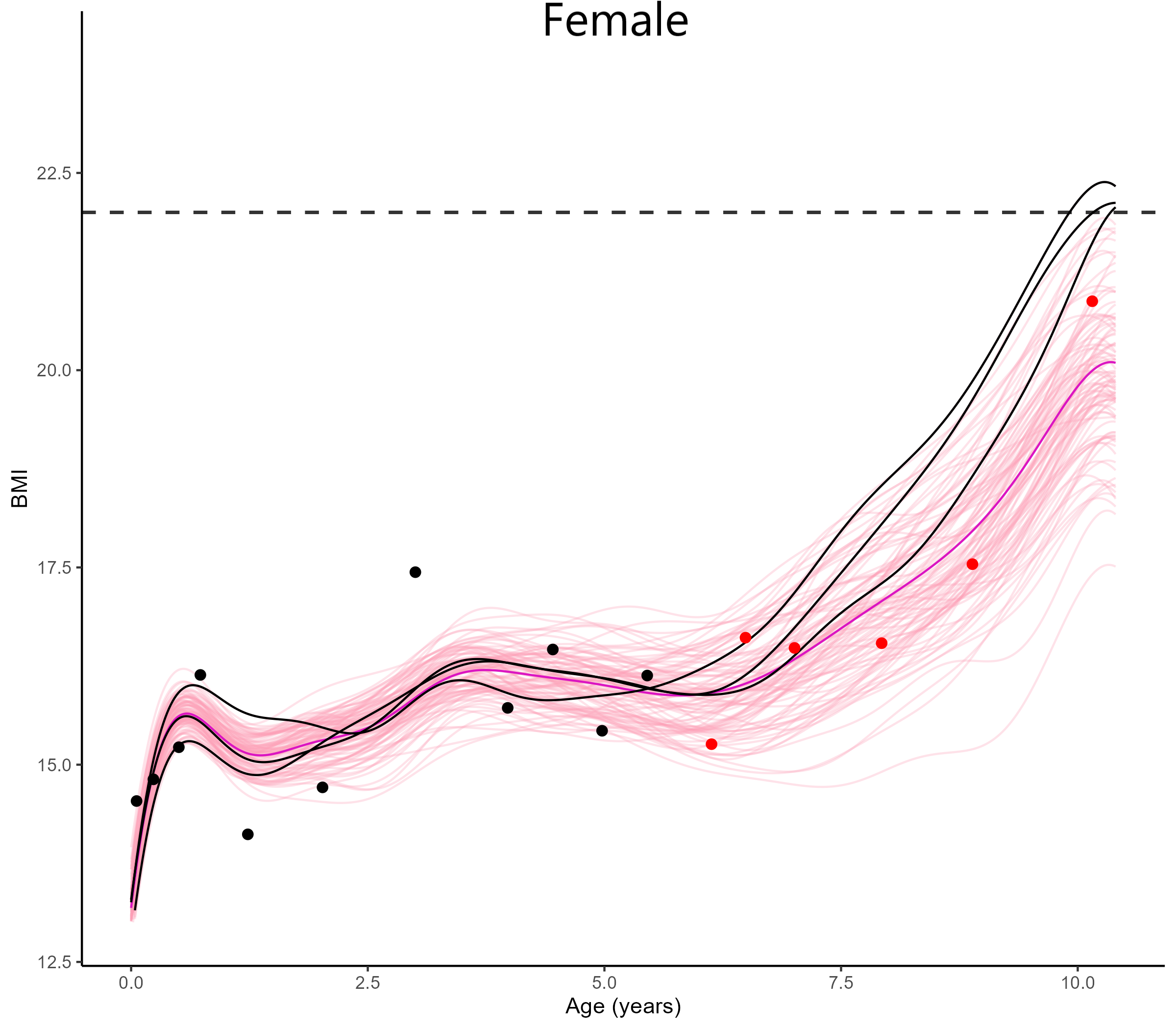}
	\caption{The overweight probability can be estimated from \textsc{MagmaClust} predictions by computing the proportion of posterior samples crossing the corresponding threshold. For male children (\textbf{left}), obesity is defined as being over $22.8 \ kg/m^2$, and 4 of the 100 samples cross this threshold in the example. For female children (\textbf{right}), obesity is defined as being over $22 \ kg/m^2$, and 3 of the 100 samples cross this threshold in the example.}
	\label{fig:risk_obesity}
\end{figure}
As an illustration, we displayed in \Cref{fig:risk_obesity} an example of this procedure for a random boy (left panel) and girl (right panel). 
In both cases, we displayed 100 predictive samples computed from the 0-6 years data for visualisation, and coloured in black the curves crossing the gender-specific overweight threshold (dashed horizontal line). 
We can notice that, although the mean trend of those predictions is below the overweight threshold, the probability of being overweight at 10 years remains non-null (4\% for the boy and 3\% for the girl). 
Such a tool provides a valuable risk quantification of undesirable events several years in advance by leveraging the well-calibrated uncertainty coming from \textsc{MagmaClust} results.

To evaluate the accuracy of our overweight probabilities and assess the applicability of such a tool in practice for overweight risk computations, we reported empirical evidence from our dataset comparing the observed and predicted overweight status for different ranges of forecasting period (from 2, 4, 6 and 8 to 10 years).
Within the 577 individuals in the testing set, only 297 (148 girls, 149 boys) had BMI data at age 10 years. 
Among them, 40 children were reported as being \emph{overweight/obese} at 10 years, and 257 were not. 
\begin{figure}
	\centering
        \includegraphics[width = \textwidth]{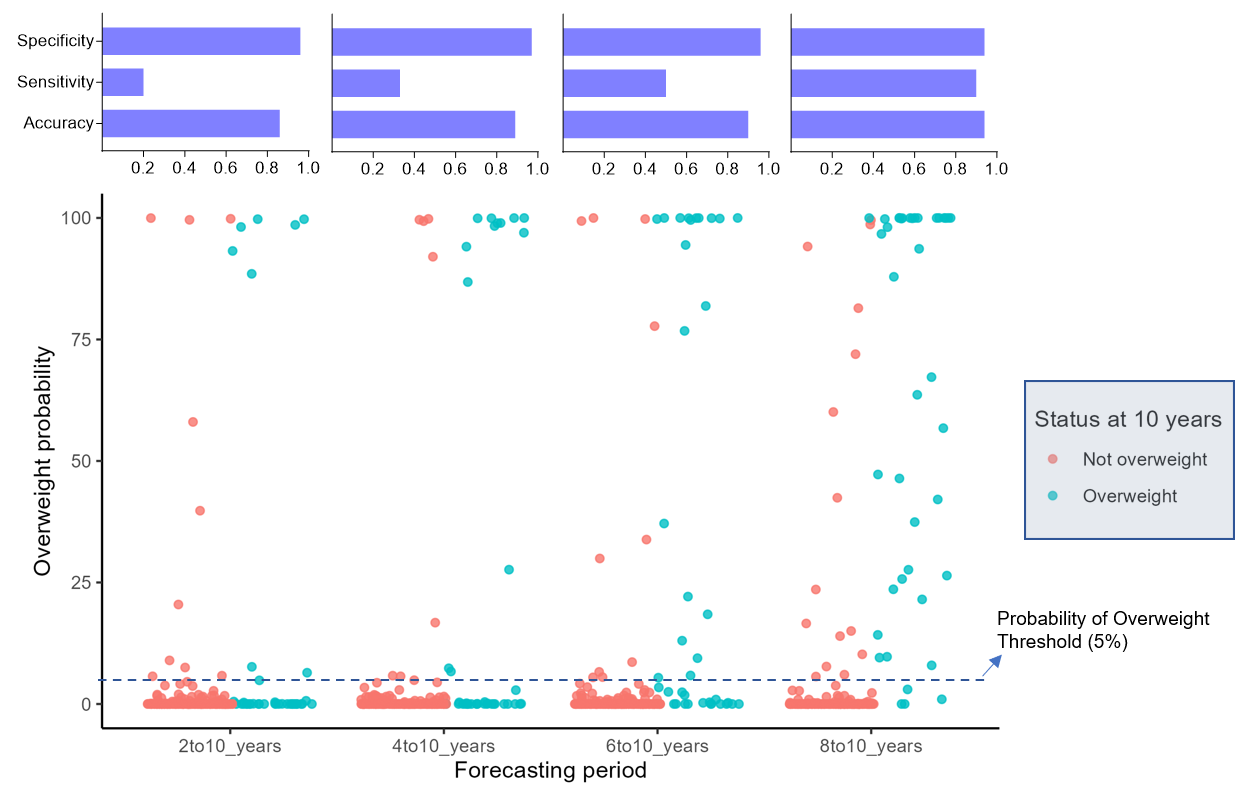}
	\caption{Visualisation of overweight probabilities estimated from \textsc{MagmaClust} predictions for different forecasting periods. Children are coloured according to their observed overweight status at 10 years (Overweight threshold is $BMI > 22$ for girls and $BMI > 22.8$ for boys). Measures of specificity, sensitivity and accuracy for a predicted overweight probability threshold of 5\% are overlaid on top of the figure.}
	\label{fig:eval_overweight_vs_truth}
\end{figure}
In \Cref{fig:eval_overweight_vs_truth}, we reported the computed probabilities of being overweight/obese for those 297 children, distinguishing them according to their observed status at 10 years. 
We can notice that, as we include more data, the identification of true overweight children seems to greatly improve. 

Although the probabilities that we compute are more informative than rough binary classifications such as \emph{at risk} vs \emph{not at risk}, by providing a quantified measure of our degree of confidence, we defined for evaluation purposes an arbitrary value of 5\% as a cut-off to compare our overweight probabilities with the observed status at 10 years.
Such a choice is rather conservative and would indicate that we aim to detect even low overweight probabilities (to implement prevention measures, for instance), though this threshold could easily be adapted depending on the clinical context and needs. 
With this convention, the overall evaluation of our overweight predictions in terms of accuracy, sensitivity and specificity are reported in \Cref{fig:eval_overweight_vs_truth}.
We can observe that the sensitivity increases largely with the number of observed points to reach 90\% of overweight children who were correctly identified 2 years in advance (from their data up to 8 years). 
Overall \textsc{MagmaClust} attributes low probabilities (0 or close) to the vast majority of children who turned out not to be overweight at 10 years, which is reassuring. 
The specificity of the method for detecting overweight/obese status at age 10 remains very high ($0.94$ to $0.96$) even for predictions starting at age 2. 


When it comes to implementing decision-making procedures, quantifying and controlling uncertainty is of paramount importance for practitioners. 
Beyond the prospective mean trends in BMI, \textsc{MagmaClust} also provides a proper uncertainty quantification.
This constitutes a precious asset to derive practical inferences from a probabilistic prediction framework.
As a final illustration, we highlighted in \Cref{fig:accounting_uncertainty} an example of what could be considered a \emph{poor} prediction on average.
Although the mean trend (of the left graph) fails to capture the true evolution of BMI for this child, we can notice that one trajectory (dashed line) among the 100 posterior samples still indicates that such a future trend was \emph{possible} albeit \emph{unexpected}. 
When collecting more data (black points), the prediction adapts its mean trend accordingly (on the right graph), and the uncertainty decreases. 
This example illustrates how uncertainty can be carefully taken into account when practitioners need to make decisions. 
Such a well-calibrated uncertainty quantification informs us of the exact degree of caution one should keep in mind before taking further action. 

\section{Discussion}

We introduced and evaluated a non-parametric and probabilistic framework named \textsc{MagmaClust}. 
This method allows us to achieve the following aims: firstly, identifying typical BMI trajectory patterns throughout childhood.
Secondly, it enables the prospective determination of individual BMI curves several years into the future, leveraging the growth history (for instance, predicting BMI until 10 years from data between birth and 2 years). 
Thirdly, \textsc{MagmaClust} provides indicators highlighting any deviations in the trajectory from the anticipated growth trend for a particular child.

The estimates of BMI trajectories from \textsc{MagmaClust} were compared to two existing methods classically used by practitioners: cubic B-Splines with fixed effects only and Jenss-Bayley's method accounting for random effects. 
We found that in terms of robustness, both \textsc{MagmaClust} and Jenss-Bayley models remain applicable regardless of the missing values proportions, as these methods account for random effects in the cohort. 
However, it was observed that before year 2, when BMI rapidly changed, \textsc{MagmaClust} accurately captured the peak region, resulting in smaller MSE.
Also, as the proportion of missing values increased, by taking off 10 to 90\% of the observed BMI, errors went from 0.90 to 2.84 in the case of \textsc{MagmaClust}, compared with 0.94 to 8.06 for Jenss-Bayley. 

Regarding BMI forecasting, only \textsc{MagmaClust} demonstrated accurate predictions up to age 10. 
The accuracy of BMI predictions remained consistent across various intervals for the forecasting period, ranging from 2-10 years to 8-10 years. 
Both for missing data reconstruction and forecasting tasks, \textsc{MagmaClust} exhibited remarkably well-calibrated uncertainty quantification in our empirical evaluations.

As an additional downstream analysis, we developed a tool to sample a large number of trajectories from the BMI probabilistic predictions to compute, at any age, the proportion of trajectories crossing the overweight thresholds. 
Such an approach provides a practical tool to assess the probability of being overweight in the future from historical growth data. 
From empirical evaluation, we reported high accuracy at all ages ($86\%$ to $94\%$) for this overweight detection procedure and a quickly increasing sensitivity, allowing us to identify from 0-2 years data $20\%$ of the overweight children at 10 years, rising up to $90\%$ when using 0-8 years data. 
The specificity was also consistently high ($94\%$ to $97\%$) over all forecasting periods.


In this work, we demonstrated the ability to prospectively predict BMI patterns at different ages and proposed tools for clinicians to detect early possible deviations from the expected trajectory.
Currently, the prospective assessment of risks for children to face obesity or overweight is generally determined by paediatricians based on the population extremes (usually the 90th centile) of BMI at any time point.
However, to the best of our knowledge, there is no reference framework for determining the most probable BMI trends over the next years. 
The method presented in this paper offers both a methodological advancement and a practical tool to monitor expected growth trends and possible deviations during childhood. 
Besides overweight issues, unusual growth trajectories may emerge even within the normal weight range.
For instance, growth slowing during periods of nutritional distress and subsequent catch-up during periods of nutritional abundance can lead to diverse health implications \citep{dulloo2006thrifty}. 
Our proposed method is responsive enough to capture this variability early on, within approximately the first year of deviation. 
By considering clusters of typical growth patterns and individual growth history to derive probabilistic predictions accounting for uncertainty, the framework provides a comprehensive approach to identify deviations from expected children's development and a practical toolbox for calibrating early intervention and risk mitigation.

In this work, we have explored arbitrary time periods (2, 4, 6 and 8 years) to illustrate and evaluate the predictive capacities of the competing methods. 
Those thresholds could be redefined based on visits to paediatricians (e.g. well-child visits and vaccination visits), as one crucial advantage of Gaussian process-based methods is to model functions over time, for which predictions can be achieved at any age.
One of the major limitations of forecasting BMI trajectories is that the future environment experienced by the child cannot be foreseen, as sudden changes in nutritional or physical activity habits can drastically influence BMI, and including such information in the model remains an open question. 

While existing literature emphasises the significance of monitoring and detailing the longitudinal growth of children to gain a more nuanced understanding of potential predispositions for adverse adult health, our work introduces a methodology for prospectively delineating growth trajectories up to the age of 10. This can be achieved with as little as a 2-year record of a child’s BMI measurements. 
Our algorithm possesses the flexibility to undergo retraining by incorporating growth trends from older age groups of children (for instance, capturing post-pubertal growth trends).
This adaptability enables the extension of predictions to older ages, providing a robust tool for ongoing assessments of childhood growth trajectories.

\section{Code availability}
The \textsc{MagmaClust} framework is implemented as an R package called \emph{MagmaClustR}, available on the CRAN, while a development version can be found on GitHub (\url{https://github.com/ArthurLeroy/MagmaClustR}). To help experiments' reproducibility, all computations, results and trained models presented in this article are stored in the following GitHub repository \url{https://github.com/ArthurLeroy/BMI_MagmaClust}.

\bibliographystyle{unsrtnat}
\bibliography{biblio}

\newpage

\section*{Supplementary}
\label{sec:supplementary}
\setcounter{figure}{0}
\renewcommand{\figurename}{Fig.}
\renewcommand{\thefigure}{S\arabic{figure}}
\setcounter{table}{0}
\renewcommand{\tablename}{Table}
\renewcommand{\thetable}{S\arabic{table}}

\begin{figure}[h]
	\centering
       \includegraphics[width = \textwidth]{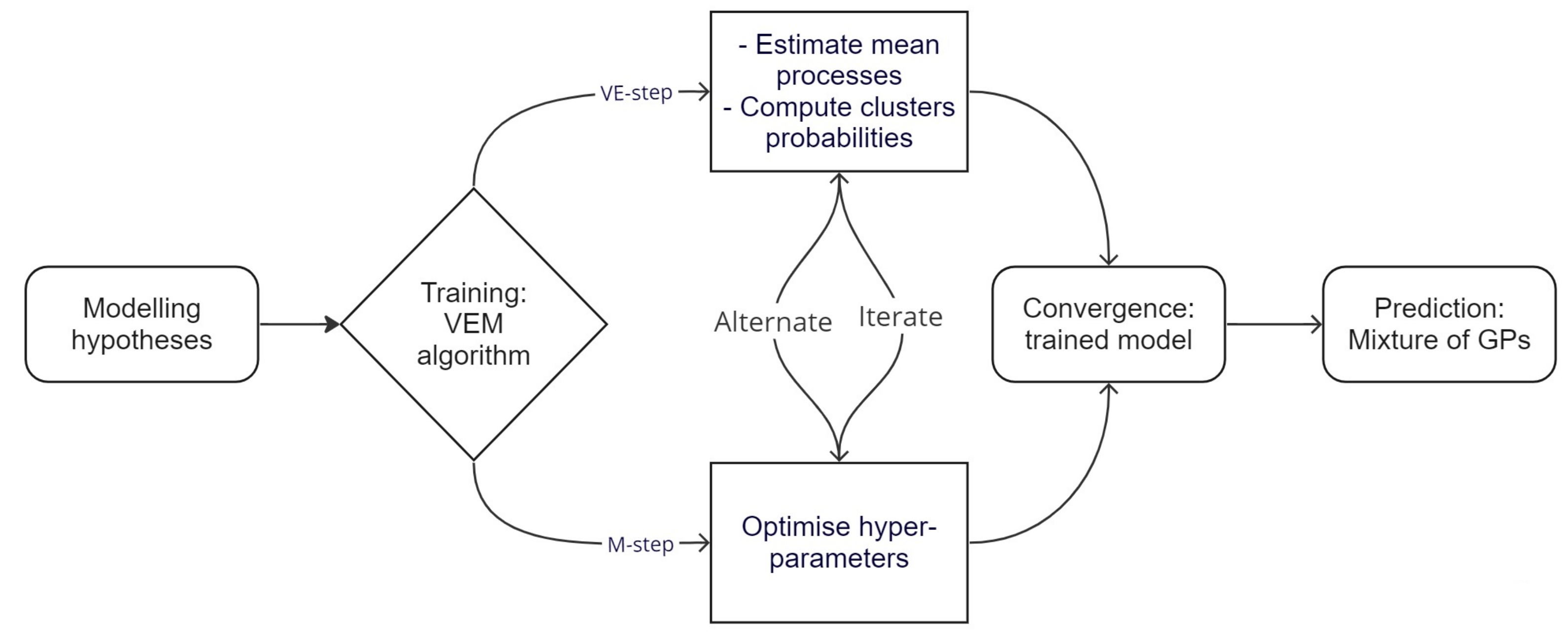}
	\caption{Flowchart summarising the main steps of the \textsc{MagmaClust} algorithm.}
	\label{fig:flowchart}
\end{figure}

\begin{figure}
	\centering
  \includegraphics[width = 0.49\textwidth]{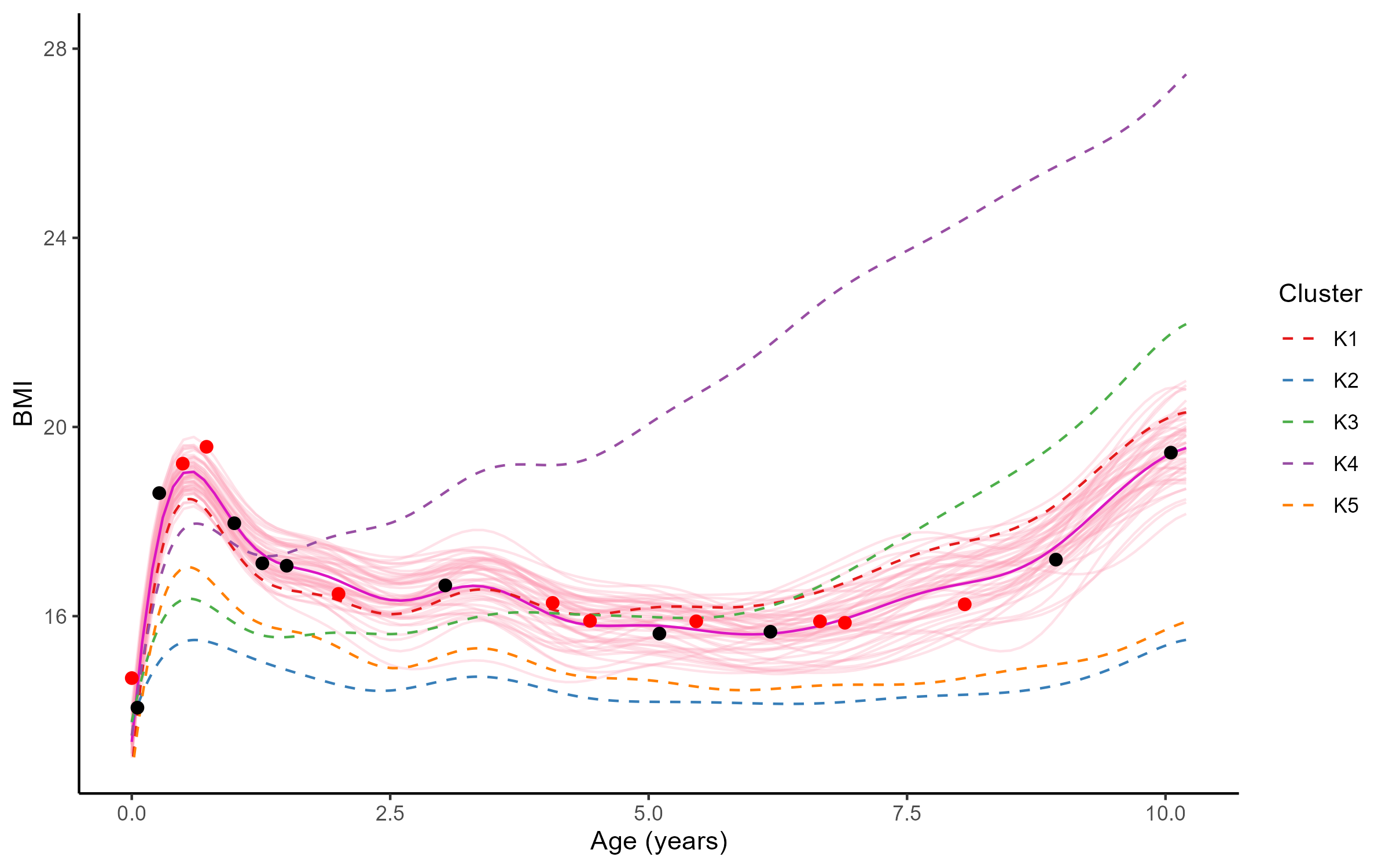}
 \includegraphics[width = 0.49\textwidth]{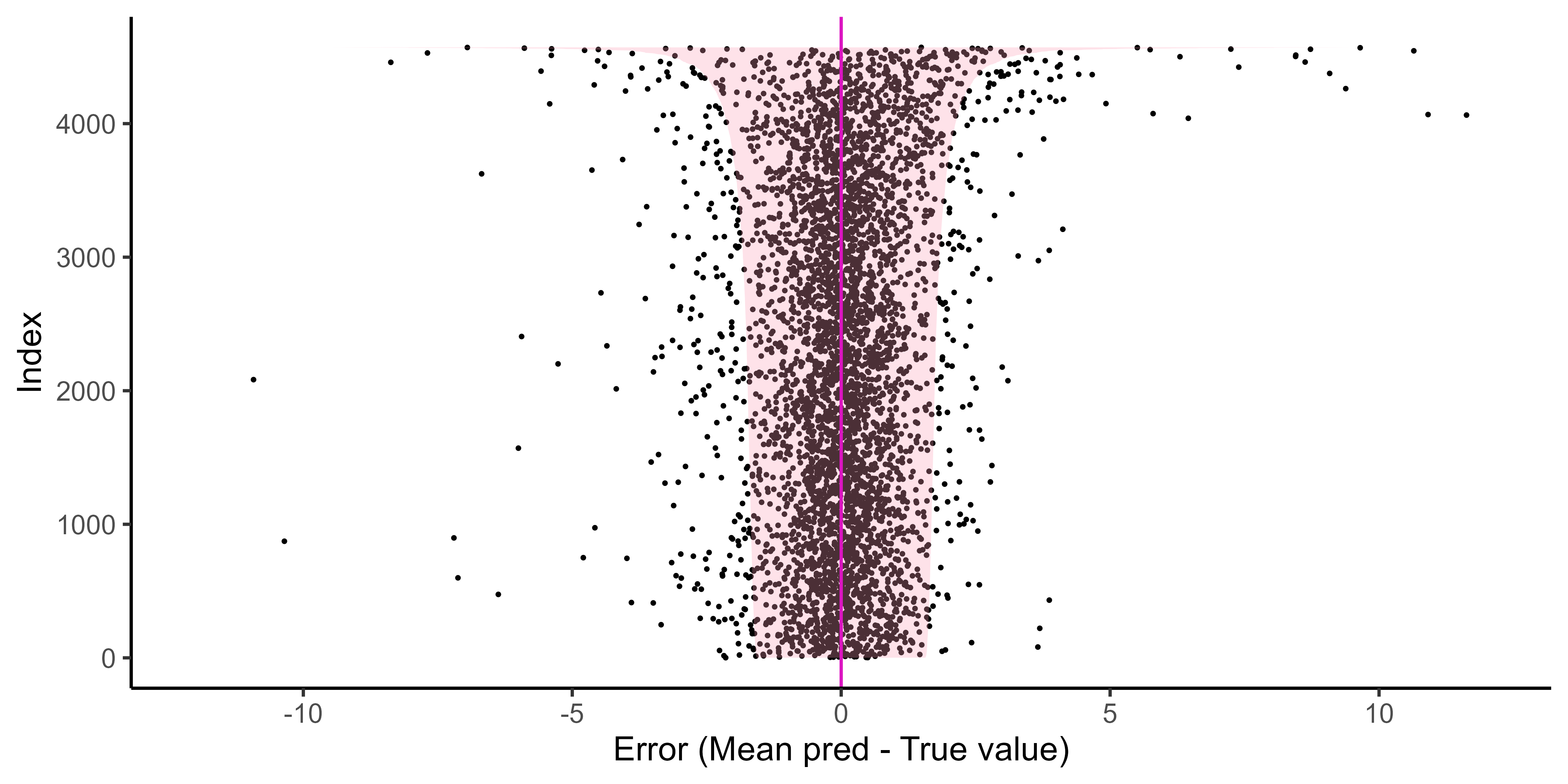}
	\caption{\textbf{Left:} Illustration of \textsc{MagmaClust} forecasts for a random illustrative individual, observed until 2 years (top), 4 years (middle), and 6 years (bottom). Observed points are in black, while testing points are in red. The purple curve represents the mean prediction, whereas the pink curves correspond to 50 samples drawn from the posterior to highlight the prediction uncertainty; \textbf{Right:} Overall uncertainty quantification of errors for forecasts across all individuals from the same times (2, 4, and 6 years). For each individual index (sorted by decreasing uncertainty on the y-axis), the 0 purple +line on the x-axis represents its posterior mean; the pink region corresponds to the associate 95\% credible interval; and the black dot is the absolute error to the true value}
	\label{fig:error_missing_uncertainty}
\end{figure}

\newpage
\begin{figure}
	\centering
        \includegraphics[width = 0.49\textwidth]{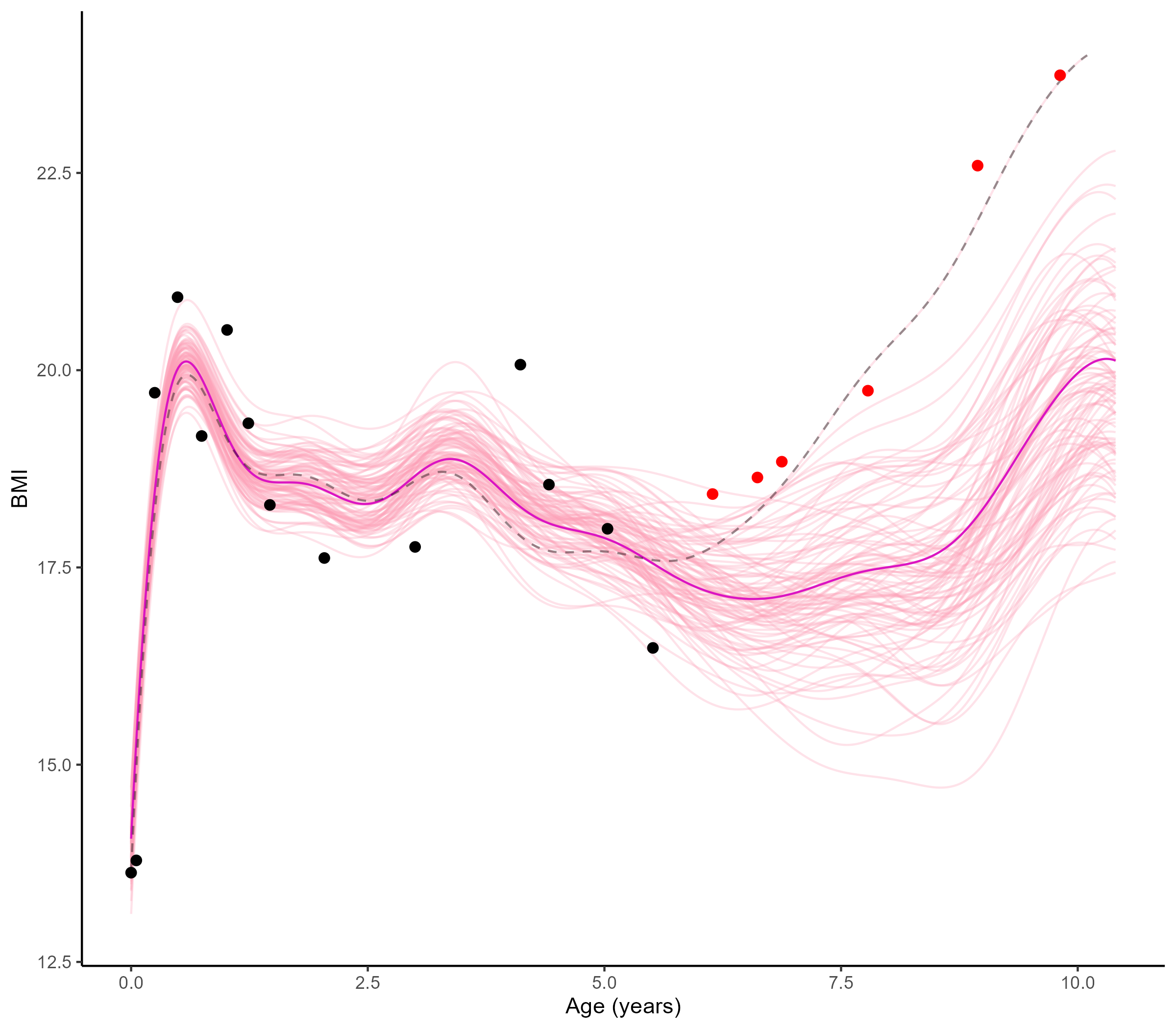}
        \includegraphics[width = 0.49\textwidth]{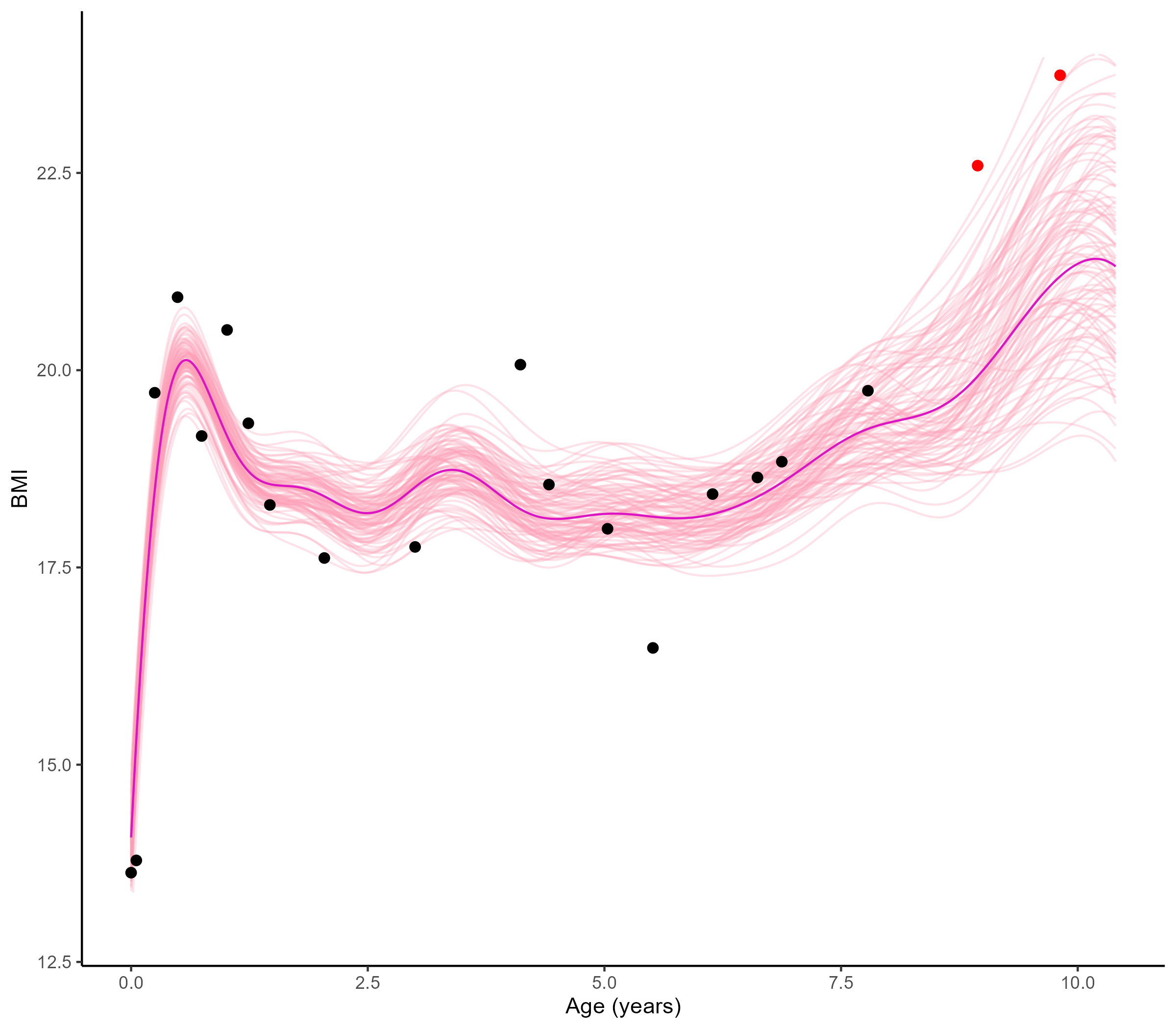}
	\caption{Example of deviation of observed BMI from \textsc{MagmaClust} predictions. Posterior sample trajectories (pink curves) are represented around the mean trend (purple curve) based on observed growth until 6 years (black dots), and the actual BMI (red dots) corresponds to low-probability trajectories. When increasing the observation range to 8 years (right panel), trajectories adapt accordingly. However, deviation from the expected growth trajectory can potentially constitute an alert for clinicians. 
 }
	\label{fig:accounting_uncertainty}
\end{figure}

\begin{table}
\centering
\begin{tabular}{|l|lll|}
\hline
Age (in month) & Weight & Height & BMI  \\ \hline
0              & 1179   & 1175   & 1175 \\
0.75           & 1038   & 1036   & 1035 \\
3              & 1024   & 1024   & 1024 \\
6              & 980    & 984    & 980  \\
9              & 941    & 942    & 941  \\
12             & 960    & 960    & 958  \\
15             & 962    & 941    & 940  \\
18             & 911    & 862    & 857  \\
24             & 925    & 892    & 891  \\
36             & 932    & 928    & 925  \\
48             & 861    & 858    & 858  \\
54             & 896    & 897    & 895  \\
60             & 873    & 873    & 872  \\
66             & 862    & 863    & 862  \\
72             & 832    & 830    & 830  \\
78             & 821    & 821    & 821  \\
84             & 863    & 864    & 863  \\
96             & 807    & 807    & 807  \\
108            & 755    & 756    & 755  \\
120            & 661    & 661    & 661  \\ \hline
\end{tabular}
\caption{Number of children with observed weight, height and BMI at each time point.
Two children did not have weight and height information at birth but had this information at other instants. Therefore, the total number of children for whom BMI trajectories were available was 1177.}
\label{tab:number_indivs_observed}
\end{table}

\end{document}